\documentclass[
 preprint,
 amsmath,amssymb,
 prl,
 superscriptaddress,
]{revtex4-1}
\usepackage{graphicx}
\usepackage{dcolumn}
\usepackage{bm}

\begin{document}
\preprint{APS/123-QED}
\thanks{A footnote to the article title}%
\title{Surface enhanced nonlinear Cherenkov radiation in one-dimensional nonlinear photonic crystal}
\author{Xiaohui Zhao}
\affiliation{Department of Physics and Astronomy, Shanghai Jiao Tong University, 800 Dongchuan Road, Shanghai 200240, China}
\author{Yuanlin Zheng}
\email{ylzheng@sjtu.edu.cn}
\affiliation{Department of Physics and Astronomy, Shanghai Jiao Tong University, 800 Dongchuan Road, Shanghai 200240, China}
\author{Huaijin Ren}
\affiliation{Institute of Applied Electronics, China Academy of Engineering Physics, Mianyang, Sichuan 621900, China}
\author{Ning An}
\affiliation{Shanghai Institute of Laser Plasma, China Academy of Engineering Physics, Shanghai, 201800 China}
\author{Xuewei Deng}
\affiliation{Laser Fusion Research Center, China Academy of Engineering Physics, Mianyang, Sichuan 621900, China}
\author{Xianfeng Chen}
\email{xfchen@sjtu.edu.cn}
\affiliation{Department of Physics and Astronomy, Shanghai Jiao Tong University, 800 Dongchuan Road, Shanghai 200240, China}
\date{\today}
\begin{abstract}
We study the configuration of efficient nonlinear Cerenkov diffraction generated from a one-dimensional nonlinear photonic crystal surface, which underlies the incorporation of both quasi-phase-matching and total internal reflection by the crystal surface. Multidirectional radiation spots with different Cerenkov angles are demonstrated experimentally, which results from different orders of reciprocal vectors. At specific angles, the incident light and total internal reflect light associating with quasi-phase-matching format completely phase-matching scheme, leading to great enhancement of harmonic efficiency.
\end{abstract}
\maketitle

\section{Introduction}

When the phase velocity of nonlinear polarization wave ($\nu_p$) exceeds that of the harmonic waves ($\nu^\prime$) in the nonlinear medium, it will emit coherent electromagnetic waves called nonlinear Cerenkov radiation (NCR) \cite{zembrod1969surface}. The Cerenkov angle is defined as $\theta=arccos(\nu_p/\nu^\prime)$, which implies the automatically longitudinal phase-matching condition between the fundamental and the radiated harmonic waves. With $\chi^{(2)}$ photonic crystals \cite{zhang2008nonlinear}, waveguides \cite{tien1970optical,li1990cerenkov} or other micro-structures assistance are introduced, Cerenkov radiation could demonstrate possibilities of a wide variety of phase-matching types and diverse patterns of spatial modulation. For instance, the phase-tuned Cerenkov-type interaction in two dimensional nonlinear photonic crystals \cite{phase-tuned}, the quasi-phase-matching associated NCR generated in waveguides or nonlinear $\chi^{(2)}$ crystals \cite{Mateos:12}, domain wall enhanced high-order NCR \cite{Chen:11,sheng2011vcerenkov,an2012cherenkov}, and so on. Such modulation mechanism has greatly expanded the NCR radiation characteristics, providing potential applications of short wavelength lasers \cite{2001Sci}, broadband frequency doubling \cite{Deng:10} and optical imaging \cite{Sheng:10,PhysRevLett.104.183901}.

In addition, the efficiency of NCR is mainly effected by the abrupt change of the second-order nonlinearity $\chi^{(2)}$, which contains not only the -1 to 1 $\chi^{(2)}$ modulation corresponding to the domain wall but also the modulation 0 to 1 corresponding to the crystal surface \cite{sheng2012role,Zhao:16}. By using sum frequency polarization wave generated by incident and internal total reflected waves \cite{ren2013enhanced}, previous studies have achieved enhanced NCR on the crystal surface \cite{ren2013surface}. Such NCR can provide good light quality and relatively high efficiency, which allows further practical applications, such as nondestructive diagnostics, harmonic conversion and ultrashort pulse characterization.

In this work, we study the behavior of NCR generated from the crystal surface and modulated by the $\chi^{(2)}$ microstructure on the surface. Using coupled wave equation, we also demonstrate the effect of reciprocal vectors of photonic crystals to the radiation angles of NCR. By utilizing the internal reflection inside the crystal boundary, the sum-frequency polarization of the incidences associated with different orders of reciprocal vectors can emit multiple NCR which exhibits $\chi^{(2)}$ spatial modulated pattern. Particularly, at specific incident angles, one can achieve degenerated NCR which leads to remarkable enhancement on the efficiency.

\section{Phenomenon and Analysis}

For simplicity, here we choose a one-dimensional (1D) periodically poled $\mathrm{LiNbO_3}$ crystal (PPLN) which provides uniform collinear reciprocal vectors along x-axis as shown in Fig. 1(a). The poling period of the sample $\Lambda=6.92 ~\mathrm{\mu m}$ and it was put on a rotation stage which can be adjusted in the y-z plane. Regarding to the calculation of the Sellmeier equation of the sample \cite{Gayer2008}, one can find the refractive index of ordinary-polarized fundamental wave is larger than that of the extraordinary-polarized second harmonic wave when the wavelength of pump is longer than 1023 nm. Consequently, it provides an anomalous-dispersion-like medium by utilizing type \uppercase\expandafter{\romannumeral1} (oo-e) SHG phase-matching interaction scheme \cite{ren2012nonlinear}. The light source we used was an optical parametric amplifier (TOPAS, Coherent Inc.) producing 80 femtosecond pulses (1000 Hz rep. rate) at the variable wavelengths from 280 nm to 2600 nm. A quarter-wave plate and a Glan-prism were set to adjust the polarization of incidences. The laser beam was loosely focused into the x-z plane of the sample and a screen located 10 cm behind the sample to receive the emitted patterns. The operating temperature was kept at 20 $^\circ$C.

\begin{figure}[htb]
\centerline{
\includegraphics[width=8.0cm]{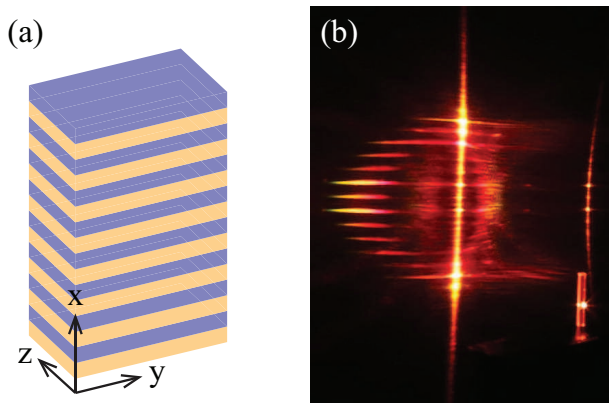}}
\caption{(a) Structure of the PPLN sample; (b) Multiple Cerenkov diffraction pattern.}
\end{figure}

When the ordinary polarized fundamental wave (FW) with a wavelength of 1250 nm injecting into the y-z plane of the sample, there was no NCR emerge owing to the phase-matching condition being not satisfied in such anomalous dispersion. As rotating the sample along x-axis, the fundamental beam would reach the x-y plane and be reflected on this surface while the incidence angle exceed the requirement of total reflection inside the sample. By adjusting the incident angle of FW, and combining the total reflection wave and reciprocal vectors, the sum-frequency polarization along the crystal surface would give birth to NCR. The far-filed image on the screen is shown in Fig. 1(b). The straight line of dots in middle is the nonlinear Raman-Nath diffraction generated from the sum-frequency of incidences and total reflection, while the right arc line belongs to the conical scattering second harmonic (SH). The fascinating phenomenon happened on the left of the sum-frequency Raman-Nath diffraction, where appeared multiple Cerenkov diffraction pattern with transverse angular dispersion lying in an arc array. Distinguished from the phenomena in bulk material, the pattern exhibits periodically spatial distribution, which denotes to the reciprocal-involved NCR by total reflection on the PPLN surface.

To analyse the distribution of SH, the coupled wave equation under paraxial and small-signal approximation was solved using the Fourier transform, and the intensity of SH $I_2$ was expressed as \cite{Zhao:16,sheng2012theoretical}:
\begin{equation}
  I_2(k_x,k_z)=\Big[\frac{k_2}{2n_2^2}\chi^{(2)}\Big]^2I_1^2L^2\mathrm{sinc}^2\Big[\Big(k_2-2k_1\mathrm{cos}\alpha-\frac{k_x^2+k_z^2}{2k_2}\Big)\frac{L}{2}\Big]|F(k_x)|^2|G(k_z)|^2,
\end{equation}
where $n_2$ is the refractive index of the SH, $I_1$ denotes the complex amplitudes of the Gaussian FW with width $a$, $L$ is the interaction distance of the nonlinear process, $\alpha$ is the incident angle of FW to $y$ axis, $k_1$ and $k_2$ are the wave vectors of the FW and SH, respectively. In the expression, $2k_1\mathrm{cos}\alpha=|\vec{k_1}+\vec{k_1^\prime}|$, where $\vec{k_1^\prime}$ is the wave vector of reflected FW. $k_x$ and $k_z$ are the components of $k_2$ in $x$ and $z$ directions. $F(k_x)=\sqrt{\frac{\pi}{2}}a\sum_{n}g_ne^{-a^2(nG_0-k_x)^2/8}$ and $G(k_z)=\sqrt{\frac{\pi}{8}}ae^{-a^2k_z^2/8}+i\frac{\sqrt{2}}{2}aD\big(\frac{ak_z}{8}\big)$ are the Fourier transform of $\chi^{(2)}$ modulated structure which are introduced by periodically reversed domains and the crystal boundary, respectively. And $g_n$ are the fourier coefficients which can expressed as:
\[ g_n=\begin{cases}
~2\mathrm{sin}(n\pi d)/(n\pi) & n\neq0\\
~2d-1 & n=0,
\end{cases}\]
where $n$ are integers. And $G_0=2\pi/\Lambda$ denotes the 0-order reciprocal vector of PPLN, $d$ is the duty ratio of domain reversal. $D\big(\frac{ak_z}{8}\big)$ denotes the Dawson function.

\begin{figure}[htb]
\centerline{
\includegraphics[width=12.0cm]{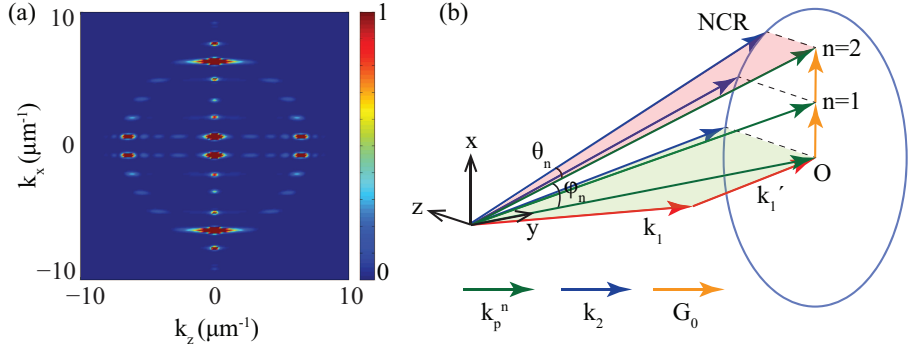}}
\caption{(a) Simulation result; (b) Phase-matching geometry.}
\end{figure}

The simulation result is shown in Fig. 2(a) under the same condition of experiment and the incident angle of FW $\alpha=20^\circ$. The distribution of SH has the similar pattern with Fig. 1(b). But the right half was total reflected in experiment. From Eq. (1), we can find that, when $k_z=0$, the SH intensity $I_2$ gathers in the direction
defined by $k_x=nG_0$ which represent the nonlinear Raman-Nath diffraction of sum-frequency. When $k_z\neq0$, SH will radiate at angles satisfied the conditions $k_x=nG_0$ and $k_2-2k_1\mathrm{cos}\alpha-\frac{k_x^2+k_z^2}{2k_2}=0$. Under the paraxial approximation, the solution of the latter is $k_x^2+k_y^2+(2k_1\mathrm{cos}\alpha)^2=k_2^2$, which is exactly the longitudinal phase-matching condition of NCR. The corresponding phase-matching geometry is shown in Fig. 2(b). The emission of 0-order NCR in the experimental pattern is in accordance with the situation in bulk material \cite{ren2013enhanced}. For high-order NCR, one should take the reciprocal vectors into consideration, which is associated with Fourier components of the $\chi^{(2)}$ modulation in terms of quasi-phase-matching (QPM) along the x-axis. The nonlinear sum-frequency polarization wave along the reflection interface (x-y plane) associating with the reciprocal vectors along the y-axis simulates an effective polarization wave which could emit the high order Cerenkov radiation. The wave vector of nonlinear polarization wave of $n$-order NCR has the form as: $\vec{k_p^n}=|\vec{k_1}+\vec{k_1^\prime}+n\vec{G_0}|$. And we can deduce the radiation angle of $n$-order NCR along $x$ axis:
\begin{equation}
\varphi_n=\mathrm{arctan}\frac{nG_0}{2k_1\mathrm{cos}\alpha},
\end{equation}
and along $z$ axis:
\begin{equation}
\theta_n=\mathrm{arccos}\frac{\sqrt{(nG_0)^2)+(2k_1\mathrm{cos}\alpha)^2}}{k_2}.
\end{equation}

\section{Experiment Results}

To experimentally demonstrate the calculated relationship of n-order NCR angles, we investigate the external angles of different order Cerenkov radiations varying with the incident wavelength, with fixed external incident angle of FW $i=30^\circ$. And with fixed wavelength of FW $\lambda=1250$ nm, we measured the relationship between the external emergence angles and the external incident angles. The experimental results, as shown in Fig. 3(a) and Fig. 3(b), respectively, demonstrate good agreement with theoretical predications. According to the phase-matching condition and the experimental results, we verify that the polarization wave is always confined along the crystal surface.

\begin{figure}[htb]
\centerline{
\includegraphics[width=12.0cm]{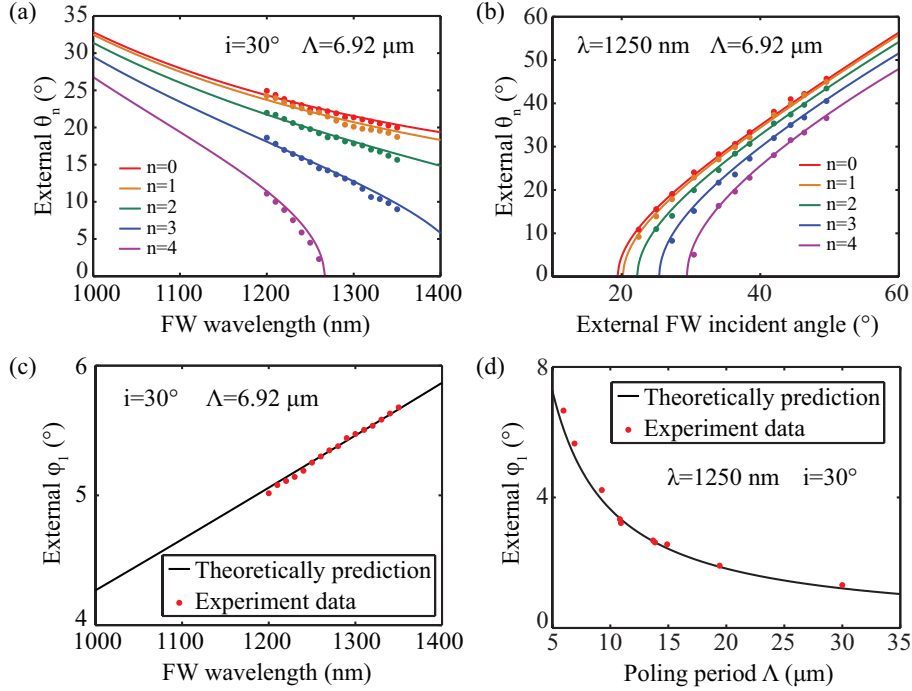}}
\caption{The external angles along z axis of different order NCR versus the incident wavelength of FW (a) and external incident angle (b). The relationship of external angles along x axis of 1-order NCR varying with the incident
wavelength (c) and the poling period of PPLN (d). Theoretical prediction (solid curves) and experimental results (signs) are in well agreement with each other.}
\end{figure}

Furthermore, we investigate the transverse distribution of the nonlinear diffraction along x axis. With external incident angle fixed as $30^\circ$, the external angles of the 1st Cerenkov radiations varied with the incident
wavelength [Fig. 3(c)]. When the wavelength of FW was fixed as 1250 nm, we draw the relationship between diffraction angles and the periods by using several samples with different poling periods, as shown in Fig. 3(d). The angular position of nonlinear Cerenkov diffraction patterns are varying associating with the variance of the samples, which implies the periodical structure are modulating the surface reflecting Cerenkov radiation. The transversely spatial distribution (along x-axis) of nonlinear Cerenkov patterns coincided with nonlinear Raman-Nath diffraction in 1D nonlinear crystal.

\begin{figure}[htb]
\centerline{
\includegraphics[width=12.0cm]{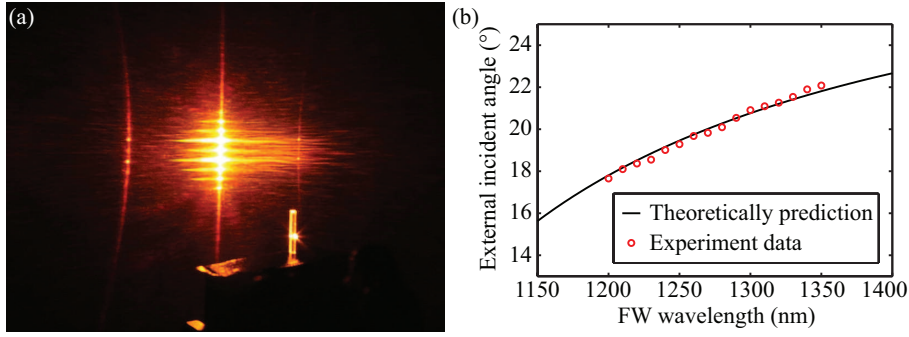}}
\caption{(a) Recorded pattern of enhanced 0-order NCR. (b) The external FW angles of enhanced 0-order NCR versus the incident wavelength of FW.}
\end{figure}

In Fig. 3(b), we note that each order of NCR has a cutoff angle respectively. At these incident angles, the emission angle of corresponding order NCR equals to 0. The phase-matching condition can be written as $\vec{k_p^n}=|\vec{k_1}+\vec{k_1^\prime}+n\vec{G_0}|=\vec{k_2}$, where the polarization wave is collinear with the Cerenkov harmonic wave and the phase-mismatch is minimized, so that each order of NCR would be greatly enhanced at these specific angles. In experiment, it's clearly that such enhancement occurred at the beginning of the NCR emergence.

We have experimentally recorded the enhanced 0-order NCR with wavelength at 1250 nm and sample period of 6.92 $\mathrm{\mu m}$, as shown in Fig.4 (a). In Fig. 4(b) we verify the dependence of the incident angle of the enhanced 0-order NCR as a function of the incident wavelength. The incident angle increased proportionally with the incident wavelength, which is consistent with theoretical analysis. So far, we have realized modulating the both diffraction pattern and the emit efficiency of Cerenkov diffraction on the crystal surface, which provide more plentiful radiation patterns compared with NCR on bulk crystal surface.

\section{Conclusion}
In summary, we theoretically and experimentally demonstrated the $\chi^{(2)}$ modulated nonlinear Cerenkov diffraction on the crystal surface. The sum-frequency polarization wave generated by incident and reflected waves is confined on the crystal surface and modulated by the periodical $\chi^{(2)}$ structure. The diffraction angle and transverse distribution of the QPM-NCR are investigated experimentally, which shows a good agreement with the theoretical calculation. In addition, with proper incident angles, the multiple diffraction would be greatly enhanced, which features the NCR emergence. It's can present more plentiful radiation patterns could be expected in other various $\chi^{(2)}$ structures, such as aperiodic, quasi-periodic, random, chirp, two-dimensional or other desirable patterns. Further, making appropriate artificial structures on the crystal surface would allow us to control the behavior of harmonic generation more efficiently.

\section*{Funding}
The National Basic Research Program 973 of China under Grant (2011CB808101); the National Natural Science Foundation of China under Grant (61125503, 61235009, 61205110, 61505189 and 11604318); the Innovative Foundation of Laser Fusion Research Center; the Presidential Foundation of the China Academy of Engineering Physics (201501023).

%

\end{document}